\documentclass[nofootinbib,preprint]{revtex4-1}
 \pdfoutput=1
 \usepackage{graphicx}
 \usepackage{amsmath}
\usepackage{amsfonts}
\usepackage{amssymb}
\usepackage{hyperref}

\def\bea{\begin{eqnarray}}
\def\eea{\end{eqnarray}}
 \def\be{\begin{equation}}
\def\ee{\end{equation}}

\begin{document}

\title{{No scale Sugra inflation with Type-I seesaw}}
 
 \author{Ila Garg}
\email[Email Address: ]{ila.garg@iitb.ac.in}
\affiliation{Department of Physics, Indian Institute of Technology Bombay, Powai, Mumbai 400 076, India}

\author{Subhendra Mohanty}
\email[Email Address: ]{mohanty@prl.res.in}
\affiliation{Theoretical Physics Division, 
Physical Research Laboratory, Ahmedabad - 380009, India}

\begin{abstract}
We show that MSSM with three right handed neutrinos incorporating a renormalizable Type-I seesaw superpotential and no-scale SURGA K\"{a}hler potential
can lead to a Starobinsky kind of inflation potential along a flat direction associated with gauge invariant 
combination of Higgs, slepton and right handed sneutrino
superfields. The inflation conditions put constraints on the Dirac Yukawa coupling
 and the Majorana masses required for the neutrino masses and also demands the tuning among the parameters. The scale of inflation is set by the mass of the heaviest right handed neutrino.  
  We also fit the neutrino data from oscillation experiments  at low scale  
  using the effective RGEs of MSSM with three right handed neutrinos. 
\end{abstract}
\maketitle

%\tableofcontents

\section{Introduction}
\label{intro}
The standard model (SM) extended with three right handed neutrinos (RHNs) is an appealing mechanism 
for explaining the light neutrino masses through seesaw mechanism 
\cite{Minkowski:1977sc,seesaw1,seesaw2,Mohapatra:1979ia,Schechter:1980,Schechter:1981cv}. 
The SM extended with supersymmetry (SUSY) \cite{Wess:1973kz} can stabilize the electroweak vacuum and help in achieving the gauge
coupling unification. Extensions of SM is also motivated by the need to explain matter anti-matter asymmetry and dark matter. 
In addition, the theory of inflation, the most successful mechanism for explaining  the large scale 
structure and anisotropy in the CMB spectrum \cite{Smoot:1992td,Bennett:2012zja,Ade:2015xua,Ade:2014xna} also needs extension of SM.
 The CMB observations \cite{Smoot:1992td,Bennett:2012zja,Ade:2015xua,Ade:2014xna},  in particular, the bounds on the tensor to scalar ratio, $r_{0.05}<0.07$ at 95\% CL and  spectral index, $n_s=0.968 \pm 0.006$, put stringent constraints on generic models of inflation  like  those arising from
the quartic potential and quadratic scalar potentials. The $R+ \alpha R^2$ inflation model of Starobinsky\cite{Starobinsky:1980te},  on the other hand, 
successfully survives these stringent constraints on $r$ and $n_s$ with the prediction that  $n_s=1 -2/N$ and $r=12/N^2 \sim 0.002-0.004$ with the minimum number of e-foldings, $N \sim 55$.

In \cite{Ellis:2013xoa}, it has been shown that the Starobinsky potential for inflation can be derived from supergravity (SURGA) with a no-scale
\cite{Cremmer:1983bf,Ellis:1984bm,Lahanas:1986uc} K\"{a}hler potential and a Wess Zumino superpotential
with specific couplings. The realization of the Starobinsky inflation in various GUT models like SO(10) \cite{Garg:2015mra,Ellis:2016ipm,Ellis:2016spb,Ellis:2014gxa} ,  SU(5) \cite{Ellis:2014dxa,Ellis:2017djk}  and flipped SU(5)xU(1)  \cite{Ellis:2017jcp} have been studied.

 In this work, we consider the minimal supersymmetric standard model (MSSM) extended with three right handed neutrinos with no-scale 
 K\"{a}hler  potential. The gauge invariant combination of a left handed sneutrino, the SM Higgs and a right handed sneutrino 
along a D-flat direction acts as inflaton. We show that this leads to Starobinsky type inflationary potential \cite{Starobinsky:1980te}. 
For realization of the Starobinsky potential in this scenario, the Yukawa coupling is related to the Majorana mass
(of the heaviest right handed neutrino) which determines the inflation dynamics. This parameter is 
fixed  by the observed amplitude of the temperature anisotropies. This value of right handed neutrino mass is then used
as an input for the see-saw neutrino mass generation. We use the RGEs of MSSM with three RHN \cite{Ibarra:2008uv} to run
down the Yukawa and eliminate the three right handed neutrinos at their respective scale to calculate the effective dimension 
five operator $\kappa$ \cite{Weinberg:1979sa}. Then, we run the effective dimension five parameter $\kappa$ using it's RGEs \cite{Babu:1993qv,Antusch:2001ck,Antusch:2001vn,Antusch:2005gp} to $M_Z$ (mass of $Z$ boson) to fit the neutrino data (mass square differences and three mixing angles). 

The inflation with right handed sneutrino has been studied in \cite{Murayama:1992ua,Murayama:1993xu,Hamaguchi:2001gw,
Ellis:2003sq,Antusch:2004hd,BasteroGil:2005gy,Kadota:2005mt,Boucenna:2014} and  snuetrino-Higgs along flat direction in 
MSSM has been studied in \cite{Allahverdi:2005mz,Allahverdi:2006iq,Antusch:2010va,Antusch:2010mv,Pallis:2011ps,Haba:2011uz,
Kim:2011ay,Khalil:2011kd, Arai:2011aa, Arai:2012em,Aulakh:2012st,Nakayama:2013nya, Kawai:2014doa, Evans:2015mta,Deen:2016zfr,
Chakravarty:2016avd}.  In particularly, \cite{Deen:2016zfr} also considers the gauge invariant combination of a left handed sneutrino,
 the SM Higgs and a right handed sneutrino along a D-flat direction, but the approach is different. The SUSY breaking scale is $\sim$ $10^{13}$ GeV which also sets the scale of inflation. The neutrino is a Dirac fermion and extremely tiny $O(10^{-12})$ third generation Yukawa coupling is considered. In our case neutrino is a Majorana particle and the neutrino Yukawa are of same order as SM Yukawa. The scale of inflation is set by the mass of third generation right handed neutrino ($10^{13}$ GeV) and SUSY breaking scale is O(50-100 TeV).

In present work, the SUSY breaking is done by adding an additional Polonyi field  which can 
acquire vacuum expectation value ($vev$) at the end of inflation. The mass of the Polonyi field 
is more than the mass of gravitino by choosing particular superpotential parameters to evade Polonyi 
problem and to get small cosmological constant \cite{Coughlan:1983ci,Ellis:1986zt,Goncharov:1984qm,Linde:2011ja,Dudas:2012wi}. 
 After the end of inflation, the decay of RHN can explain the asymmetry 
through leptogenesis which can be converted into the Baryon asymmetry through Sphaleron process \cite{Fukugita:1986hr}.

In Section $\bf{2}$, we present our inflation model. In Section $\bf{3}$, we briefly discuss about 
reheating and  SUSY breaking in our model. 
In Section $\bf{4}$, we give a benchmark input satisfying the inflationary
conditions along with the neutrino oscillation data. Then, we conclude with a brief discussion.

\section{Inflation along D-flat LNH direction} 
The superpotential considered for inflation contains the terms sufficient to give neutrino masses via type-I seesaw mechanism. 
The relevant superpotential is given as,
 \begin{equation}
    W=  Y_{\nu}^{ij} L_i H_u N_j+ \frac{1}{2} M_{N}^{jj} N_jN_j +\mu H_u H_d
\label{WSIMSSM}\end{equation}
 First term is the Dirac term and the second term is Majorana mass term for right handed neutrinos (N). 
 Here, $i,j$ represents the number of generations of fermions. Also, $Y_{\nu}$ is a complex matrix and $M_{N}$ is real diagonal. 
 Here, $H_d$ is another Higgs doublet required for anomaly cancellation in MSSM.
The K\"{a}hler potential is assumed to be of  the general form {\footnote {The moduli field $T$ can be stabilised by adding extra terms 
 $((T+T^*-1)^4+d(T-T^*)^4)/\Lambda^2$ inside the log term of the K\"{a}hler potential \cite{Ellis:2013nxa}. 
 The inflationary potential has  a flat direction and Starobinsky form along Re(T)=1 and Im(T)=0 in Planck units (see Fig. (2) of \cite{Ellis:2013nxa} ).}} given as,
\be 
K= -3\ln \left(T+T^*- \sum_i \,k_i \phi_i \phi_i^*  \right)\ee
where we will choose  the constants $k_i=1/3$ for the fields $L_1$, $N_3$, $H_u$ whose linear combination will constitute the inflaton and we will choose $k_i \ll 1$ for all other fields
(we are working in the units where Planck mass scale, $M_P$=$(8\pi G)^{-1}$=1). Therefore for the inflation components the K\"{a}hler potential is of the 
no-scale SURGA form, given as,
\be K= -3\ln \left(T+T^*-\frac{1}{3}(|L_1|^2+|N_3|^2+|H_u|^2  ) \right)\ee
and for all other fields the K\"{a}hler potential goes to the canonical form
\be
K = \delta_{ij} \phi_i \phi_j^*
\ee

 The corresponding potential is given as,
  \be V= e^{G} \biggr[\frac{\partial G}{\partial \phi^m} K^m_{n^*}\frac{\partial G}{\partial \phi_{n}^*}-3\biggr] +\frac{1}{2}D^aD^a
 \label{potential}\ee
 where,
 \be G=K+\ln W+\ln W^* \,\,\,\,\,\,\ D^a = g^a \left(K_m ({\cal T}^a)^m_n \phi_n \right) \ee
and $K^m_{n^*}$ is the inverse of K\"{a}hler metric $K_{m}^{n^*}$. Here m, n runs over the number of fields,
 $g$ is the gauge coupling and ${\cal T}'s$ are the generators of each gauge group in SM. 
 The kinetic term is given as $K_{m}^{n^*} \partial{\phi^m} \partial{\phi_{n}^*}$.

 The scalar part of  $SU(2)_L$ Higgs and lepton doublets appearing in the Yukawa term during inflation (setting charged component to zero) can be written as,
 \be L_i= \begin{pmatrix} \tilde\nu_i \\ 0\end{pmatrix}; \quad H_u =\begin{pmatrix}0 \\ h_u\end{pmatrix}; \quad H_d =\begin{pmatrix}h_d \\ 0\end{pmatrix}.\ee
 The D-flat direction for the gauge invariant combination $LHN$ ($D^a$ = 0) is given by, 
\be \sum_i |\tilde\nu_i|^2=|h_u|^2  \ee

 The right handed neutrino is gauge singlet and doesn't contribute to D-term.  Also, we have three generations of neutrinos and the freedom to 
 choose any generation of sneutrino
for our inflaton. We chose the 3rd generation right handed neutrino assuming the normal hierarchy of neutrino masses and first generation left handed neutrino. 
 We will assume that all other scalar vevs are zero during the course of inflation to make sure that the fields are non zero in $LNH$ flat direction only 
 and are zero in any other flat direction. So, we consider 
the fields $\tilde N_3$, $\tilde
\nu_1 $ and the Higgs field $H_u$ parametrized in terms of
a D-flat direction associated with the gauge invariant $LHN$  and $NN$ terms in the superpotential,
\begin{equation}
\tilde{N_3} = \tilde{\nu_1} = h_u = \phi \label{inflaton},
\end{equation} 
to be the inflaton. 
The superpotential and K\"{a}hler potential
relevant for inflation is given as,   
\bea W= Y^{13}_{\nu} \phi^3 + M^{33}_{N} \frac{\phi^2}{2} \,\,;\,\,\,\,\,\,\,
K=-3\ln(T+T^*-|\phi|^2)\eea
After simplifying, the potential and  the K.E. has the following form,
\bea V= \frac{1}{(1-|\phi|^2)^2} \biggr| \frac{\partial W}{\partial \phi }\biggr|^2 \label{PF} \,\,;\,\,\,\,\,\,\,
L_{K.E.}=  \frac{3}{(1-|\phi|^2)^2} |\partial^{\mu} \phi|^2 \eea
Here, we have assumed that the non-perturbative Planck scale dynamics fixes the value of $T=T^*=\frac{1}{2}$.
After fixing the $vev$ for $T$ the kinetic
terms of $T$ can be neglected.
 To get the canonical K.E. terms, we need to redefine our fields in terms of
 the new field $\chi$ as,
 \be \phi= \tanh \frac{\chi}{\sqrt{3}} \ee
 Now, for $\chi$ = $x+iy$, the complex part of $\chi$ is fixed to zero during inflation since it has a mass greater than the Hubble rate \cite{Ellis:2013xoa} .
 Then considering the  condition $Y_{\nu}^{13}$= -$M_N^{33}$, the potential for the real part of $\chi$ field looks like,
\be V = {M_N^{33}}^2(1-e^{-\frac{2 x}{\sqrt{3}}})^2 \ee
This is the Starobinsky kind of potential for inflation. Here the scale of the inflation is set by the mass of the heaviest right handed neutrino mass, $M_N^{33}$.   
All other fields have conventional $m^2 \phi^2$ potentials which  subdominant  ( $m^2 \ll M_{33}^2 $) and steep compared to the inflation potential and so that the fields stole at zero during the slow roll of the inflation without destabilizing the inflation potential.
The slow roll parameters for this potential are given by,
\be \eta=-\frac{8 e^{\frac{-2 x }{\sqrt{3}}} \left(1-2e^{\frac{-2 x }{\sqrt{3}}}\right)}{3
   \left(1-e^{-\frac{2 x }{\sqrt{3}}}\right)^2}; \quad  \epsilon= \frac{8 e^{-\frac{4 x }{\sqrt{3}}}}{3 \left(1-e^{-\frac{2 x
   }{\sqrt{3}}}\right)^2}\,.
   \label{SR}
   \ee
   When $\eta$ $\approx$ 1, inflation ends and this corresponds to field value, $x^{end}$ $\approx$ .5.
   The required number of $N_{e-folds}$=55 to have sufficient inflation gives the initial field
   value of $x$ $\approx$ 4.35. The power spectrum for scalar perturbation,
   \be P_R = \frac{V}{24\pi^2 \epsilon} = \frac{{M_N^{33}}^2 \sinh ^4\left(\frac{x}{\sqrt{3}}\right)}{4 \pi ^2}\,, \ee
   requires the value of $M_N^{33}$ = $7.87 \times 10^{-6}$ in Planck units  for the central value of $P_R = 2.2  \times 10^{-9}$ given by Planck data \cite{Ade:2015xua}.
     The spectral index $n_s$ =0 .964 and tensor to scalar perturbation ratio, r =0.002 for $N_{e-folds}$=55. However, 
     the deviation from the condition $Y_{\nu}^{13}$= -$M_N^{33}$,  even at forth decimal place leads to large deviation from the observational data \cite{Ellis:2013xoa}.
The soft mass terms of the inflaton fields will also contribute to the inflation potential.
But, we take SUSY breaking scale O(TeV) small as compared to $M_N^{33}$. 
We discuss about SUSY breaking in the next section.  

\section{Reheating and SUSY breaking}

After the end of inflation, the hot big bang conditions can be restored when energy stored in the inflaton
is converted into a thermal bath of the MSSM
degrees of freedom. The time required to
thermalise the inflaton energy and the resulting reheat
temperature $T_{rh}$ depends on the post inflationary dynamics of the LHN flat direction \cite{Garg:2015aga}.  
 In \cite{Garg:2015aga}, the post inflationary dynamics of the LHN flat direction has been studied in $U(1)_R$ $\times$ $U(1)_{B-L}$.
 Here, we haven't extended our gauge sector, but the procedure follows the same. 
 Due to the Yukawa couplings and gauge coupling of inflaton fields (however, $N$ has only Yukawa coupling), the inflaton energy
is likely to decay rapidly (within one Hubble time) through the so-called instant preheating
mechanism \cite{Kofman:1994rk,Kofman:1997yn} to radiation bath of MSSM degrees of freedom.  So in present scenario,
the estimate for the maximum reheating temperature  is given as \cite{Garg:2015aga},

\bea T_{rh} = \left(\frac{30}{ \pi^2 g_{*}}\right)^{\frac{1}{4}} V_0^{\frac{1}{4}} \sim 10^{13} GeV \eea
Here, $g_{*}$=228.75 is the MSSM degrees of freedom. 
In this model, one can explain the baryon asymmetry, $n_{B}/n_{\gamma}$, through leptogenesis.
The large reheat temperature ensures the thermal production of the heavy neutrinos $N_3$ whose decay could produce leptogenesis. 
However any existing lepton asymmetry will be washed out by the Higgs-neutrino scattering upto the temperature $T_{washout}=10^{12}$ GeV. 
The lepton asymmetry arising from the decay of the lighter right handed neutrinos $N_1$ and $N_2$ whose masses are less than $10^{12}$ GeV 
can in principle generate leptogenesis which can be converted to baryogenesis by spahlerons.

Also, an important point to be made is that such a large reheating
temperature can produce relativistic
populations of gravitinos, which are dangerous if
the lifetimes of gravitinos are larger than the time of nucleosynthesis,
$\tau_N$ $\sim$ 1 sec, and their decay after nucleosynthesis will overcome the entire universe.
This is famous ``gravitino problem'' and the general solution of this problem is to have the graviton
mass sufficiently large so that it decay before nucleosynthesis \cite{Kawasaki:2005}.
\bea \tau_{grav} \sim 10^5 sec
\left(\frac{1 TeV}{m_{3/2}} \right)^3 \ll \tau_{N} \sim 1 sec \eea
So, the viability of this model requires supersymmetry breaking scale O(TeV)
with gravitino mass $m_{3/2}$ $\sim$ 50 TeV. However, with gravitino mass O(50) TeV, there is an upper bound on the reheating temperature $\sim ~O(10^9)$ GeV \cite{Kawasaki:2017bqm}.
For $T_{rh}$ $>$ $10^9$ GeV, the lightest supersymmetric particle (LSP) produced from the decay of gravitino may over-close the universe. The possible way out is that the  overproduced LSP should decay through small R-partity violation before BBN \cite{Barbier:2004ez,Nakayama:2016gvg}.

The minimal superpotential and K\"{a}hler potential responsible for inflation can not give rise to SUSY breaking.
The SUSY can be broken by adding a Polonyi field, $S$ \cite{Linde:2011ja,Dudas:2012wi} and adding the following terms,
 \bea K(S,\bar S)= S\bar S+\frac{(S \bar S )^2}{\Lambda^2}  \,\,;\,\,\,\,\,\,\,
W(S)=M^2 S+ \Delta \eea 
to the  K\"{a}hler potential and Superpotential respectively.
   
 The term, $(S \bar S)^2/\Lambda^2$ with $\Lambda$ $\ll$1 and the fine tuning of the constant,
$\Delta$ help in the strong stabilization of the Polonyi field and 
fixing the vanishingly small cosmological constant $\sim$ $10^{-120}$.
Also, the cosmological Polonyi problem \cite{Coughlan:1983ci,Ellis:1986zt,Goncharov:1984qm} can be solved with the condition,
 \bea m^2_S \gg m^2_{3/2}, \eea
 so that, it decay into gravitinos. This can be achieved with $\Delta$ $\neq$ 0 and for $\Lambda$ $\ll$1 and the potential minimum $V_{min} \approx -3\Delta^2 + M^4$ with 
$S_{min} \approx \Delta \Lambda^2 /2M^2$. Therefore, for $M^2$ $\approx$ $\sqrt{3}\Delta$, the cosmological
constant is very small $\sim$ $10^{-120}$. This gives $S_{min} \approx \Lambda^2/2\sqrt{3}$.
The gravitino mass is given as
\bea m^2_{3/2}=e^{G}= \frac{1}{(T+T^*)^3}ln(S\bar S+\frac{(S \bar S )^2}{\Lambda^2})|W(S)|^2.\eea
So, at the minimum of the potential the garvitino and Polonyi field masses (in Planck units) are obtained as,
\bea m^2_{3/2} = \Delta^2,\quad\quad m^2_{S}= \frac{12\Delta^2}{\Lambda^2}= \frac{12m^2_{3/2}}{\Lambda^2} \gg m^2_{3/2}, \eea
respectively.  For $ \Lambda$ $\sim$ $10^{-2}$ and $\Delta \simeq 2\times 10^{-15}$,
we obtain $m_{3/2}$ $\sim$ 50 TeV and $m_S \sim$  500 TeV.  The inflation potential after SUSY breaking takes the form
\be
V={M_N^{33}}^2(1-e^{-\frac{2 x}{\sqrt{3}}})^2 + m_{3/2}^2 \left( \tanh \frac{x}{\sqrt{3}} \right)^2
\label{V1}
\ee
With $m_{3/2}$ = 50 ${\rm TeV}$ $\ll  M_{33}$ the second term is subdominant compared to the first term in (\ref{V1}). 
In addition, the slow roll parameters are thus same as that given in  (\ref{SR}). The SUSY breaking terms, therefore, does not destabilize the inflaton potential.
 
 \section{Example fit to neutrino oscillation data}
The superpotential given in Eqn. (\ref{WSIMSSM}) is also responsible for neutrino masses  through Type-I seesaw. 
The required ingredients for Type-I seesaw are $Y^{ij}_\nu$ and $M^{jj}_N$. The mass matrix, $M$ for the $\nu$ and $N$ from eqn. (\ref{WSIMSSM}) can be written as,
\bea M=  \begin{pmatrix}
  0& M_D \\
M_D^T & M_N 
\end{pmatrix} \eea
Here, $M_D=Y_{\nu}v$ with $v$=246 GeV, the SM $vev$. 
For $M_N$ $\gg$ $M_D$, the masses of right handed neutrinos are given by $M_{N}$ and the tiny masses of left handed neutrinos are given as, 
\bea M_{\nu}= \frac{1}{2}M_{D}^T M_{N}^{-1}M_{D}.\eea
However, the three right handed neutrinos masses are not degenerate, so we use the RGEs of Yukawa of MSSM with three RHN \cite{Ibarra:2008uv}  and eliminates the three right handed neutrinos at their thresholds to calculate the effective dimension five operator, $\kappa$ \cite{Weinberg:1979sa}.  Then, we run this effective dimension five parameter, $\kappa$ using RGEs \cite{Babu:1993qv,Antusch:2001ck,Antusch:2001vn,Antusch:2005gp}  to $M_Z$  and calculate the neutrino mass matrix (3 $\times$ 3) at $M_Z$. It is given as,
 \be M_{\nu}= v^2 \kappa(M_{Z})\ee
where $\kappa= \frac{1}{2}Y_\nu M_N^{-1} Y_{\nu}$.
Using this $M_{\nu}$, we fit  the Neutrino oscillation data along with satisfying the inflation conditions. 
 For this, we use the standard parametrization of the PMNS matrix given by,
\begin{center}\bea U_\nu = \begin{pmatrix}
  c_{12}c_{13} & s_{12}c_{13} & s_{13} e^{-i\delta} \\
 -c_{23}s_{12}-s_{23}s_{13}c_{12}e^{i\delta} & c_{23}c_{12}-s_{23}s_{13}s_{12}e^{i\delta} & s_{23}c_{13} \\
 s_{23}s_{12}-c_{23}s_{13}c_{12}e^{i\delta} & -s_{23}c_{12}-c_{23}s_{13}s_{12}e^{i\delta} & c_{23}c_{13}
\end{pmatrix} P\eea \end{center}
where $ c_{ij} = cos\theta_{ij} \,\, , \,\, s_{ij} = sin\theta_{ij}$ and the phase matrix 
$P = \textrm{diag}\, (1,\,e^{i\phi_1}, \, e^{i(\phi_2 + \delta)})$  contains the Majorana phases.

Also, we have considered the following constraints on neutrino masses from various experiments. 
 \begin{itemize}
 \item The Planck 2015 results put an upper limit on the sum 
  of active light neutrino masses to be \cite{Ade:2015xua}
   \be\Sigma \, = \, m_1 + m_2 + m_3 < 0.23 \, \textrm{eV} .\ee

The global analysis \cite{Capozzi:2016rtj,Esteban:2016qun} 
of neutrino oscillation measurements with three 
light active neutrinos give the oscillation parameters in their $3\sigma$ range,  for normal hierarchy (NH)  for which $m_3 > m_2 >m_1$:  
\item Mass squared differences
\bea \label{osc1} \Delta m_{21}^2/10^{-5} \textrm{eV}^2 \, = \, (7.03 \rightarrow 8.09)\nonumber\\
 \Delta m_{31}^2/10^{-3} \textrm{eV}^2 \, = \, (2.407 \rightarrow 2.643) 
    \eea

  \item Mixing angles
   \bea  \textrm{sin}^2\theta_{12} \, = \, (0.271 \rightarrow 0.345)  \nonumber\\
    \textrm{sin}^2\theta_{23} = (0.385 \rightarrow 0.635) \nonumber\\
     \textrm{sin}^2\theta_{13} = (0.01934 \rightarrow 0.02392) 
\eea
\item Dirac Phase
\bea 
\delta_{PMNS}= (0 \rightarrow 2\pi ) 
\eea

\end{itemize}

 We randomly choose the sixteen parameters of complex $3 \times 3$ matrix $Y_{\nu}$ and two masses of heavy right 
 handed neutrinos while the component $Y^{13}_{\nu}$ is determined in terms of $M_N^{33}$ = $7.87 \times 10^{-6} $ $M_{P}$ to
  satisfy the inflation condition given in the inflation section. 

 We fit  to the neutrino oscillation data within $3\sigma$ given by experiment using a downhill simplex method \cite{Press:1996}. 
 One example input for $Y_{\nu}$ and $M_N$( GeV) is given in eqn. (\ref{ynuMN}) and the corresponding output is given in Table \ref{benchmark1}. 
 \bea Y_{\nu}&=&\begin{pmatrix}
5.21 \times 10^{-6} +5.59 \times 10^{-6} i&
   1.43 \times 10^{-5} - 3.49 \times 10^{-6} i&
  -7.87 \times 10^{-6} 
 \\
 
   7.72 \times 10^{-4} + 6.98 \times 10^{-6} i &
  -3.62 \times 10^{-5} + 7.12 \times 10^{-4} i&
   -4.14 \times 10^{-4} + 2.92 \times 10^{-5} i\\

  -2.45 \times 10^{-2} -8.26 \times 10^{-5} i&
   -2.98 \times 10^{-2}  +3.49 \times 10^{-4} i&
  -0.11 + 4.72 \times 10^{-2} i \end{pmatrix} , \nonumber\\&&
  M_N= \begin{pmatrix}
  1.84 \times 10^5& 0&0\\
  0& 1.29 \times 10^9&0\\
 0& 0&  1.91 \times 10^{13}\end{pmatrix} .\label{ynuMN}\eea 

\begin{table}[ht]
\caption{output for one benchmark point for $Y_{\nu}$
 and $M_N$ given in eqn. (\ref{ynuMN})}  
{\begin{tabular}{lll} 
\hline\noalign{\smallskip}
 {\mbox {Parameter} }&{\mbox {Value}}\\
 \hline
   $(m^2_{12})/10^{-5}(eV)^{2}$&            7.9261\\
   $(m^2_{23})/10^{-3}(eV)^{2}$&            2.4071\\
          $\sin^2\theta^L_{12}$&            0.2838\\
          $\sin^2\theta^L_{23}$&            0.4180\\
          $\sin^2\theta^L_{13}$&            0.0237\\
                $\delta_{PMNS}$&            3.0245\\
               $\phi_1,\phi_2$ &            4.7266,            6.2218\\
 \hline
\end{tabular} \label{benchmark1}}
\end{table}
 
 We can see from the example input the smallness of off-diagonal elements 
 of $Y_{\nu}$ can be easily achieved while having the diagonal entries O(0.1). So, we can achieve inflation and fit the  
 neutrino oscillation data with very realistic Yukawa coupling in this scenario.

\section{Discussions}
In this paper, we have shown how a renormalizable superpotential responsible for neutrino masses 
in MSSM +3 RHN with a no-scale type K\"{a}hler potential can lead to Starobinsky type inflationary potential. 
The scale of the inflation is set by the mass of the heaviest right handed neutrino  and is essentially independent
of the supersymmetry-breaking parameters. 
The inflation constraints the seesaw parameter values (Yukawa and Majorana mass), but the freedom of choosing the generation makes this constraint very light
 and fitting the neutrino oscillation data can be achieved very easily with realistic Yukawa couplings. 
The high reheat temperature, $T_{rh}$ $\sim$  $10^{13}$ GeV requires a gravitino mass 
 $\sim$ 50 TeV to remain consistent with nucleosynthesis. 
This model predicts a tensor-scalar ratio, $r=0.002$ and any observation in the near future above this 
value will rule out this model. The predicted masses of the heavy right handed neutrinos can be tested in
future in constructing models of leptogenesis by heavy neutrino decays. Also, SUGRA models with non-canonical K\"{a}hler potential, like the no-scale model
 discussed in this paper, predict relations between observables like the scale of inflation, SUSY breaking scale \cite{Covi:2008cn}
 and the non-gaussianity \cite{Hetz:2016ics} which may be testable in future observations.

\section{Acknowledgements}
Ila Garg would like to thank Mansi Dhuria and Tanushree Basak for useful discussions.  
\bibliographystyle{unsrt}
\bibliography{references}
\end{document}